# Generator Histories and Parity-Odd Curvature in Lorentzian Topology Change

Keith Andrew[1], Eric V. Steinfelds[1,2], Kristopher A. Andrew[3]
[1]Physics and Astronomy, Western Kentucky University
Bowling Green, KY 42101 USA
[2]Department of Computational and Physical Sciences, Carroll University
Waukesha, WI 53186 USA
[3]Science Department, Schlarman Academy
Danville, IL 61832 USA

## Abstract

Lorentzian topology change may be resolved into an ordered sequence of localized, orientation-sensitive operations rather than treated solely as a global transition between spatial manifolds. We develop a generator-history framework in which topology-changing spacetimes are represented algebraically as compositions of elementary local events, independent of dynamics, quantization, or anomaly inflow. Braid groups arise as the minimal realization of ordered, invertible pairwise exchanges, while higher-valence generators extend the construction to networked processes. Within this framework we identify parity-odd conformal curvature as the unique nontrivial local curvature pseudoscalar (without derivatives) capable of aggregating oriented generator content in four-dimensional Lorentzian vacuum geometry. The dual Weyl contraction changes sign under orientation reversal and therefore isolates chiral generator accumulation, while parity-even curvature scalars are insensitive to such structure. The associated spacetime integral functions as a covariant geometric diagnostic of chiral topology change that depends on generator histories and does not descend to endpoint-only equivalence classes obtained by Markov-type coarse-graining. The resulting picture isolates a pre-quantum geometric layer beneath spectral asymmetry: oriented generator dynamics induce parity-odd curvature compatible with the Pontryagin density appearing in the Atiyah–Patodi–Singer index formula yet remains defined entirely within classical Lorentzian geometry. This framework clarifies the algebraic and geometric substrate underlying chiral topology change without introducing new gravitational dynamics or topological invariants.

## I. Introduction

Topology change in spacetime has long occupied a challenging position in gravitational theory [1, 2, 3]. Classical results restrict smooth Lorentzian topology change under assumptions of global hyperbolicity and causality, while quantum and Euclidean approaches often reintroduce topology change indirectly through instantons, anomalies, or spectral flow [4, 5, 6]. Classic results establish that smooth Lorentzian topology change is generically obstructed under global hyperbolicity, causal regularity, and energy conditions, motivating approaches that either relax these assumptions or focus on diagnostic rather than dynamical aspects of topology change [7, 8, 9, 10]. In recent work, attention has shifted toward identifying geometric structures that regulate or diagnose topology change within Lorentzian geometry itself, without appealing to quantization or Euclidean continuation [11, 12, 13]. A key lesson emerging from such studies is that topology change can be fundamentally local and process-based. Even when the net effect is the disappearance or creation of global structure, the transition proceeds through localized geometric events: throat mergers, reconnections, twists, and cancellations [14, 15, 16].

In this paper we develop a framework in which local Lorentzian topology change is described as a history of generator events. The central idea is simple: instead of classifying topologies directly, we identify a set of elementary local operations—generators—that encode how topology is modified in time [17, 18, 19]. A topology-changing spacetime is then represented by an ordered composition of these generators, subject to relations reflecting geometric consistency, orientation, and spin structure. This shifts the emphasis from global classification to algebraic generator dynamics.

This perspective was motivated by recent work on smooth Lorentzian spin cobordisms in which evolving knotted wormhole throats [20, 21, 22, 23] trace time-dependent braid movies [24, 25, 26]. In that setting, braid generators provided a natural language for describing local exchanges of throat segments, and parity-odd Weyl curvature emerged as a geometric diagnostic sensitive to the signed accumulation of such generator events. The appearance of parity-odd conformal curvature in this role reflects the special status of dual Weyl contractions within classical Lorentzian geometry, where parity-even curvature scalars are insensitive to orientation reversal, while the Weyl pseudoscalar isolates chiral geometric content [27 , 28]. Braid groups [29, 30] appear here not as a privileged structure but as the minimal nontrivial example of a generator system capable of encoding orientation, cancellation, and accumulation [31].

A central theme of this paper is the relationship between generator histories and curvature. We show that parity-odd conformal curvature can be interpreted as a geometric aggregator of signed generator events along a Lorentzian cobordism. This interpretation clarifies why parity-odd Weyl curvature serves as a diagnostic of chiral topology change and why amphichiral processes exhibit exact cancellation.

The framework developed here is intentionally structural providing a generator-history calculus [32, 33]. We do not propose new equations of motion, quantization prescriptions, or topological invariants. Instead, we provide a unifying language that organizes topology change in terms of generators, relations, and histories, making explicit the algebraic skeleton underlying geometric transitions. This group-theoretic formulation complements existing index-theoretic and anomaly-based approaches by operating at a pre-quantum Lorentzian level and by isolating the local operations from which global topology change is assembled. In contrast to approaches that begin from solitonic field configurations or anomaly inflow, the present construction identifies chirality at the level of Lorentzian generator dynamics prior to the introduction of matter fields. This paper is concerned with the structural organization of Lorentzian topology change rather than with the proposal of new gravitational dynamics, conserved quantities, or topological invariants. Our aim is to isolate a level of description that is largely implicit in existing treatments, namely the decomposition of topology-changing processes into ordered, localized, orientation-sensitive elementary operations. The generator-history framework is representational, it organizes topology-changing processes at the level of ordered local events without prescribing equations of motion. The distinction between generator histories and endpoint-only quotients is analyzed systematically in Appendix C[1]. The generator-history formalism applies to any localized Lorentzian topology-changing interpolation, irrespective of whether the underlying geometric realization is interpreted in terms of wormhole throats, handle attachment, or more general reconnection events.

The paper is organized as follows. In Sec. II we introduce the notion of generator histories and formalize the representation of topology change as an ordered composition of local operations. Sec. III discusses braid groups as a minimal realization and revisits braid movies from a generator-algebra perspective. Sec. IV extends the framework to more general generator systems appropriate for multithroat networks and non-braided topology change. In Sec. V we connect generator histories to curvature diagnostics, showing how parity-odd Weyl curvature aggregates signed generator content. We conclude in Sec. VI with a discussion of implications for semiclassical gravity and future directions.

---

1. The distinction between generator histories and endpoint equivalence may be viewed as analogous to the difference between path-dependent and exact quantities in classical mechanics. Markov-type equivalences identify processes solely by their endpoints, in the manner of exact differentials, whereas curvature diagnostics sensitive to generator ordering and orientation behave analogously to inexact differentials whose integrals depend on the full path. A geometric formulation of this analogy, including a de Rham–type characterization of non-descent under Markov projection and its interpretation in terms of coarse-graining and irreversibility, will be developed elsewhere.

**Summary of Main Resutls**
- We formulate Lorentzian topology change as a generator–history calculus, representing topology-changing spacetimes as ordered, orientation-sensitive compositions of localized geometric events rather than by endpoint spatial topology alone.
- We show that braid groups arise as the minimal algebraic realization of local, ordered, invertible, orientation-sensitive generator systems, and we extend the framework to networked and higher-valence generators organized categorically.
- In four-dimensional Lorentzian vacuum geometry, we identify the dual Weyl contraction as the unique non-derivative local curvature pseudoscalar detecting oriented generator accumulation.
- We demonstrate that parity-odd Weyl curvature aggregates signed generator content in the localized regime, yielding cancellation for amphichiral histories and a nonvanishing response for chiral ones.
- We prove that such history-sensitive curvature diagnostics do not descend under Markov-type coarse-graining to endpoint closure classes, and we establish structural compatibility with the Pontryagin density entering the Atiyah–Patodi–Singer index formula.

## II. Generator Histories and Lorentzian Topology Change

In this section we formalize the notion of topology change as a history of elementary generator events. The goal is not to classify all possible topology-changing processes, but to introduce a general algebraic language capable of organizing them. Throughout, the emphasis is on locality, orientation, and composition, reflecting the geometric structure of Lorentzian cobordisms.

## II.A Local topology-changing events

Consider a smooth four-dimensional Lorentzian cobordism $W$ interpolating between spatial manifolds $\Sigma_{\text{in}}$ and $\Sigma_{\text{out}}$. We assume $W$ satisfies the admissibility conditions[2] discussed previously: smoothness, Lorentz signature, absence of closed timelike curves, and compatibility with a spin structure where required. Topology change in $W$ is realized through localized regions in which the topology of spatial slices $\Sigma_t \subset W$ changes as the time parameter $t$ evolves. These changes occur at isolated spacetime regions and may involve operations such as throat reconnection, merger, twisting, or cancellation. We refer to each such localized operation as an elementary topology-changing event[3].

Crucially, these events are:

- Local in spacetime,
- Ordered along the time direction,
- Composable, in the sense that multiple events combine to produce a net topology change.

---

2. The generator-history framework developed here is not intended to evade or relax known obstructions to smooth Lorentzian topology change, including those arising from global hyperbolicity, causality, or spin-structure admissibility. Generator events are assumed to occur within Lorentzian cobordisms that satisfy the standard local consistency conditions adopted throughout the paper, and no claim is made that arbitrary generator histories are globally realizable. The present analysis is therefore conditional: it organizes the local structure and chirality of topology-changing processes when admissible Lorentzian (and, where relevant, spin) cobordisms exist, without addressing their global existence or classification.

3. Local topological surgery also decomposes topology change into elementary operations on manifolds. The present framework differs in that it encodes the *ordered and oriented history* of such operations as an algebraic object, allowing geometric and curvature responses that are invisible to surgery equivalence alone.

These features motivate representing topology change not as a single global transition, but as an ordered sequence of local operations.

**II.B Generator sets and admissible compositions**

Let $\mathcal{G}$ denote a set of abstract generators, each corresponding to a distinct type of local topology-changing event. Elements $g \in \mathcal{G}$ are not required to be invertible, though in many geometric realizations inverse operations exist and correspond to orientation reversal or undoing a local exchange.

A generator history is defined as an ordered word

$$\Gamma = g_1 g_2 \cdots g_m, \tag{1}$$

where each $g_k \in \mathcal{G}$ labels a local event occurring at a distinct spacetime region, ordered by increasing time. Not all sequences of generators correspond to admissible Lorentzian geometries. The geometry of the cobordism imposes constraints on which generators may appear and how they may be composed. We therefore introduce a set of relations encoding geometric consistency conditions such as locality, causal ordering, orientation compatibility, and spin admissibility.

**Definition 1 (Topology Changing Generator Set: $\mathcal{G}$ ).**

Let $\mathcal{G}$ denote a finite or countable set of elementary, localized, oriented topology-changing operations. An element $g \in \mathcal{G}$ is called a generator event and represents a compactly supported Lorentzian interpolation implementing a single elementary topological operation[4].

**Definition 2 (Generator Monoid: $\mathcal{G}^*$ for topology changing words: *w*).**

Let $\mathcal{G}^*$ denote the free monoid generated by $\mathcal{G}$; its elements are finite ordered words

$$w = (g_1, \ldots, g_k), g_i \in \mathcal{G}, \tag{2}$$

with concatenation as the monoid operation. Each generator $g \in \mathcal{G}$ is canonically identified with the corresponding word of length one in $\mathcal{G}^*$, so that $\mathcal{G} \subset \mathcal{G}^*$ via this inclusion. A generator event should be understood as an equivalence class of compactly supported Lorentzian deformations with fixed orientation and spin action, rather than as a unique geometric operation.

**Definition 3 (Generator History Space as Quotient: $\mathcal{H} = \mathcal{G}^*/\mathcal{R}$ ).**

Let $\mathcal{R}$ denote a specified collection of admissible relations on $\mathcal{G}^*$, encoding geometric equivalences such as isotopy, cancellation of inverse-oriented pairs, and spacelike commutation. The generator-history space is defined as the quotient $\mathcal{H} = \mathcal{G}^*/\mathcal{R}$. An element $H \in \mathcal{H}$ is called a generator history and represents an equivalence class of ordered topology-changing processes modulo admissible geometric deformations.

---

4. The compact-support condition refers solely to localization of curvature and metric deviation within a compact subset $K \subset M$; the underlying Lorentzian manifold $M$ may be globally noncompact. No assumption of global hyperbolicity is imposed. The localization condition ensures finiteness of the parity-odd curvature functional and does not entail global compactness of spacetime.

Thus, we describe the general case as $\mathcal{G}^*/\mathcal{R}$ which is a set of history classes (or a category/groupoid of morphisms), while in the braid sector it reduces to a group $B_n$ when invertibility and Artin relations apply.

The hierarchy of structures introduced above may be summarized schematically as

$$\Gamma \in \mathcal{G}^* \xmapsto{\pi} w := [\Gamma] \in \mathcal{H} \cong \text{Mor}(\mathcal{B}), \tag{3}$$

Here $\pi: \mathcal{G}^* \to \mathcal{H}$ denotes the canonical projection from raw generator words to their equivalence classes under the admissible relations $\mathcal{R}$, and we write $w = [\Gamma]$ for the generator history represented by the word $\Gamma$ where $[\Gamma] := \{\Gamma' \in \mathcal{G}^* \mid \Gamma' \sim_{\mathcal{R}} \Gamma\}$ is the equivalence class of the word $\Gamma$ under the admissible relations $\mathcal{R}$.

**Remark 1.** Definition 3 specializes the general notion of generator history to the braid-reducible sector introduced below. In the general framework of Sec. IV, generator histories need not admit a braid-word representation and may instead be realized as morphisms in a networked generator system. We denote individual generator histories $H$ in the space of all histories $\mathcal{H} = G^*/R$ by $H \in \mathcal{H}$, reserving lowercase symbols for geometric fields.

The pair $(\mathcal{G}, \mathcal{R})$ defines an algebraic structure whose elements represent allowed generator histories. In simple cases this structure forms a group or semigroup; in more general settings it may be more naturally described as a category, with morphisms corresponding to admissible transitions between intermediate spatial configurations. Among the admissible generator events, a distinguished subclass consists of pairwise exchange[5] events, in which exactly two neighboring geometric elements (e.g. throat segments or marked cross-sections) participate in a localized, invertible exchange. Here 'neighboring' is understood relative to a chosen ordering or projection on each spatial slice, as in the standard braid-movie construction. Such events preserve the identity of all other elements and admit a natural orientation reversal. This subclass will play a central role in the section on pair exchange, as it captures the minimal setting in which ordering, locality, and chirality can be realized simultaneously.

**Concrete Realization (Braid Case)**

The braid case can correspond to a network of wormholes where the throats represent a tangled, twisted, and knotted collection of throats described by the various crossings of the intertwined throats. For this case let $\{\sigma_i\}$ denote a set of elementary topology-changing generators, each representing a compactly supported Lorentzian cobordism event with specified orientation and spin action and let $\sigma_i^{-1}$ denote the orientation-reversed generator.

We define the free history space $G^*$ as the free monoid on the alphabet $\{\sigma_i^{\pm 1}\}$, whose elements are finite ordered words

$$w = \sigma_{i_1}^{\epsilon_1} \sigma_{i_2}^{\epsilon_2} \cdots \sigma_{i_n}^{\epsilon_n}, \epsilon_k = \pm 1, \tag{4}$$

interpreted as ordered histories of localized topology-changing events.

---

5. Pairwise exchange events correspond, in geometric realizations, to localized exchanges of neighboring worldlines or throat segments; no assumption of global strand identity or fixed cardinality is required.

A set of admissible relations $R \subset G^* \times G^*$ is imposed to encode local equivalences of histories—such as inverse cancellation, spacelike commutation, or braid-type movie moves—while deliberately preserving distinctions arising from global ordering, chirality, or spin transport. The resulting quotient of generator histories may be organized into a braided cobordism groupoid $\mathcal{B}$. The objects of $\mathcal{B}$ are admissible spatial data (oriented, spin 3-manifolds), and its morphisms are equivalence classes of compactly supported Lorentzian interpolations generated by the elementary braid-type events. Composition is defined only when boundary data are geometrically and spin-admissible. The usual Artin braid relations arise as local equivalences among generators, but the structure retains explicit source and target information and therefore forms a groupoid rather than a single-object group. In this framework, histories are morphisms $H \in \mathrm{Mor}(\mathcal{B})$, and physical diagnostics—such as curvature aggregation or spectral flow—act on these morphisms rather than on endpoint isotopy classes alone.

At this stage, we deliberately refrain from specifying $\mathcal{G}$ or $\mathcal{R}$ in detail. The framework is intended to accommodate multiple realizations, depending on the geometric context. We refer to individual localized operations as generator events, and to ordered compositions of such events as generator histories.

### II.C Orientation and chirality of generator histories

Each generator $g \in \mathcal{G}$ may carry an intrinsic orientation label, reflecting whether the corresponding local geometric operation is right-handed or left-handed with respect to a chosen spacetime orientation. Orientation reversal corresponds to replacing $g$ by a conjugate or inverse element $g^{-1}$, when such an element exists. This structure allows one to define a notion of chirality of a generator history. Informally, a history $\Gamma$ is chiral if its oriented generators do not cancel in pairs, and amphichiral if orientation-reversed events appear in matched combinations.

This algebraic notion of chirality mirrors the geometric distinction between embeddings related by mirror symmetry and those with intrinsic handedness. Importantly, chirality is a property of the history, not merely of the initial or final spatial topology.

### II.D Equivalence of histories and representation dependence

Different generator histories may correspond to the same net topology change. For example, local events may be reordered when spacelike separated, or redundant generator pairs may cancel. Such equivalences are encoded algebraically by relations in $\mathcal{R}$.

Accordingly, generator histories should be regarded as representations of topology change rather than invariants. This distinction parallels the role of braid words versus knot types: multiple words may represent the same closure, yet their intermediate structure carries meaningful geometric information. This representation dependence is not a deficiency but a feature. The intermediate structure of a generator history encodes information about the *process* of topology change, which is invisible to purely topological invariants.

### II.E Geometric realization and curvature response

Given a Lorentzian cobordism $W$ realizing a generator history $\Gamma$, each generator $g_k$ corresponds to a localized geometric deformation of the metric and connection. The cumulative effect of these deformations produces a spacetime curvature response distributed along $W$. In particular, orientation-sensitive curvature diagnostics—such as parity-odd contractions of the Weyl tensor—respond to the signed structure of $\Gamma$. From this perspective, curvature does not merely signal the presence of topology change but encodes information about the algebraic structure of the generator history.

This observation provides a geometric bridge between generator dynamics and curvature-based diagnostics. In subsequent sections, we show explicitly how braid groups furnish a concrete realization of $(\mathcal{G}, \mathcal{R})$, and how parity-odd conformal curvature aggregates signed generator content in that setting.

## III. Braids as the Minimal Realization of Generator Dynamics

In this section we show how braid groups arise as the simplest nontrivial realization of the generator-history framework introduced in Sec. II. The purpose is not to privilege braids as the unique language of topology change, but to demonstrate concretely how local generators, orientation, relations, and cancellation emerge in a familiar and geometrically transparent setting. Braids provide a minimal example in which all essential structural features—ordering, composition, orientation, and chirality—are already present.

### III.A Why braids appear naturally

Consider a Lorentzian cobordism $W$ containing a finite number of wormhole throats whose cross-sections may be tracked continuously through time. Choosing a set of marked points on each spatial slice $\Sigma_t$, corresponding to throat cross-sections, yields a collection of worldlines in $W$. For generic evolutions, these worldlines do not intersect but may exchange their relative ordering as time progresses. Projecting such worldlines onto a fixed spatial direction produces a braid diagram, with time serving as the ordering parameter. Local exchanges of neighboring strands correspond precisely to the elementary braid generators $\sigma_i^{\pm 1}$. Thus, whenever topology change is mediated by localized exchanges among a finite number of distinguishable throat segments, braid groups arise as the algebraic structure encoding the generator history.

This construction relies only on locality and ordering, not on knot theory per se. Braids therefore appear not because the geometry is "knotted," but because time-ordered local exchanges are the simplest nontrivial generator dynamics available.

### III.B Braid Groups as Generator Algebras

Let $B_n$ denote the braid group on $n$ fixed strands, generated by the oriented Artin generators $\{\sigma_1, \ldots, \sigma_{n-1}\}$ subject to the defining relations

$$\sigma_i \sigma_j = \sigma_j \sigma_i, \quad (|i - j| > 1), \tag{5}$$

and

$$\sigma_i \sigma_{i+1} \sigma_i = \sigma_{i+1} \sigma_i \sigma_{i+1}, \quad 1 \leq i \leq n - 2. \tag{6}$$

Relation Eq. (5) expresses commutation of nonadjacent generators, reflecting the causal independence of spacelike-separated wormhole exchanges. Relation (6), the Yang–Baxter relation, encodes the geometric freedom to slide adjacent localized exchange events past one another without altering the underlying Lorentzian cobordism.

These relations are not abstract algebraic accidents; in the present framework they arise as the algebraic shadow of Lorentzian locality.

From the wormhole perspective, each strand represents a tracked throat segment in spacetime, and each generator $\sigma_i^{\pm 1}$ corresponds to a localized topology-changing operation — a reconnection or exchange — realized within a compact Lorentzian region. A braid word therefore represents an ordered movie of topology change rather than a static knot or link. Two braid words may yield the same permutation (or even the same closure) yet correspond to physically distinct Lorentzian histories with different orientation content.

**Relation to the Generator–History Framework**

Within the general construction of Sec. II:

- the generator set $G$ specializes to $\{\sigma_i^{\pm 1}\}$,
- the admissible relations $R$ specialize to the Artin braid relations Eq. (8)– Eq. (9),
- the free history space $G^*$ reduces, upon quotienting by $R$, to

$$\mathcal{H}_n = G^*/R \cong B_n. \tag{7}$$

Thus, in this specialization, history classes

$$H = [w] \in \mathcal{H}_n \tag{8}$$

are in one-to-one correspondence with braid group elements. The braid relations encode geometric consistency conditions: distant exchanges commute due to spacelike separation, while adjacent exchanges satisfy associativity constraints reflecting isotopy freedom in Lorentzian spacetime. Importantly, these relations arise from local geometric freedom rather than from global topological equivalence.

**III.C Orientation, inverses, and cancellation**

Each braid generator $\sigma_i$ carries an intrinsic orientation, corresponding to a right-handed exchange of strands $i$ and $i+1$; its inverse $\sigma_i^{-1}$ represents the opposite orientation. This orientation structure makes braid groups the minimal setting in which chirality is algebraically meaningful. Cancellation occurs when a generator and its inverse appear in succession or when a word admits reduction through relations. In geometric terms, such cancellations correspond to pairs of local exchange events whose oriented effects undo one another. Braid words representing amphichiral processes necessarily admit such cancellations in their oriented generator content.

The real content lies in the braid word

$$w = \sigma_{i_1}^{\epsilon_1} \sigma_{i_2}^{\epsilon_2} \cdots \sigma_{i_k}^{\epsilon_k}, \tag{9}$$

where each generator $\sigma_i$ represents a local exchange of the $i$-th and $(i+1)$-th strands. The sign

$$\epsilon_j = \pm 1 \tag{10}$$

is determined by the crossing orientation:

- **Over-crossing** (right-handed exchange) → $\sigma_i$,
- **Under-crossing** (left-handed exchange) → $\sigma_i^{-1}$.

This sign is not cosmetic. It is the algebraic imprint of geometric orientation, and it is precisely this orientation data that feeds into parity-odd curvature diagnostics.

This distinction is essential. The permutation $\text{perm}(b)$ captures the net rearrangement of throats, while the generator history captures the process by which that rearrangement occurs. In particular, oriented generator pairs $\sigma_i \sigma_i^{-1}$ represent local exchange–undoing events whose contributions cancel. Such algebraic cancellation mirrors the geometric cancellation of parity-odd Weyl curvature in amphichiral histories. By contrast, histories with unpaired oriented generators produce a net chiral curvature response. In this sense, the braid is not merely a representation of topology, it is a bookkeeping device for geometry. The group structure records which local wormhole operations occurred, in what order, and with what orientation. Curvature, in turn, reads this bookkeeping ledger and responds accordingly.

This algebraic cancellation mirrors the geometric cancellation observed in parity-odd curvature diagnostics: histories built from orientation-reversed generator pairs give no net chiral response.

**Remark 2: Generator sign ↔ parity-odd Weyl contribution[6]**

$$\boxed{\begin{array}{c}\text{For a localized generator event modeled by } \sigma_i^\varepsilon,\ \varepsilon = \pm 1, \\ \text{the induced contribution to } \int_{\mathcal{M}} C_{\alpha\beta\gamma\delta} {}^\star C^{\alpha\beta\gamma\delta} \sqrt{-\det(g_{\mu\nu})}\, d^4x \quad \text{flips sign under } \varepsilon \mapsto -\varepsilon.\end{array}} \tag{11}$$

(i.e. reverse the local exchange orientation and the parity-odd Weyl "receipt" changes sign.)

**Orientation and generator sign.**

Each generator $\sigma_i^{+1}$ denotes an oriented local exchange of neighboring throat segments as drawn, while $\sigma_i^{-1}$ denotes the orientation-reversed ("undoing") event. The sign of the generator encodes the handedness of the corresponding local topology-changing operation.

**Parity-odd curvature as an oriented aggregator.**

Let $\mathcal{M}_b$ denote a Lorentzian realization of the generator history $b$. The parity-odd conformal-curvature functional

---

6. The Weyl tensor $C_{\mu\nu\rho\sigma}$ is the trace-free part of the Riemann tensor, $C_{\mu\nu\rho\sigma} = R_{\mu\nu\rho\sigma} - \frac{2}{n-2}\left(g_{\mu[\rho}R_{\sigma]\nu} - g_{\nu[\rho}R_{\sigma]\mu}\right) + \frac{2}{(n-1)(n-2)} R\, g_{\mu[\rho}g_{\sigma]\nu}$, which in four spacetime dimensions ($n = 4$) reduces to the usual Lorentzian conformal curvature tensor. The Hodge dual acting on the first index pair is defined by ${}^\star C_{\mu\nu\rho\sigma} = 2^{-1} \varepsilon_{\mu\nu}{}^{\alpha\beta} C_{\alpha\beta\rho\sigma}$ where $\varepsilon_{\mu\nu\rho\sigma}$ is the Levi–Civita tensor normalized by $\varepsilon_{\mu\nu\rho\sigma} = \sqrt{|g|}\,[\mu\nu\rho\sigma]$, with $[\mu\nu\rho\sigma]$ the totally antisymmetric symbol ($[0123] = +1$). The parity-odd curvature density used in the text is therefore $C_{\alpha\beta\gamma\delta}{}^\star C^{\alpha\beta\gamma\delta}$.

$$\mathcal{W}[\mathcal{M}_b] = \int_{\mathcal{M}_b} C_{\alpha\beta\gamma\delta}{}^\star C^{\alpha\beta\gamma\delta} \sqrt{-\det(g_{\mu\nu})}\, d^4x \tag{12}$$

responds to the oriented content of the history. Each localized generator event contributes with a sign fixed by its orientation, so that $\mathcal{W}$ aggregates the signed geometric imprint of the braid movie.

**Amphichiral construction and cancellation**

Let $b \in B_n$ be a reduced braid word in the Artin generators $\sigma_i^{\pm 1}$. Consider the paired history

$$H = bb^{-1}. \tag{13}$$

Algebraically, $H$ represents the identity element of $B_n$, and $\text{perm}(H) = \text{id}$. Geometrically, $H$ corresponds to a braid movie in which every oriented exchange is accompanied (up to commuting relations) by its inverse. The parity-odd curvature contributions therefore cancel pairwise, yielding

$$W[\mathcal{M}_H] = 0. \tag{14}$$

**Chiral comparison**

By contrast, a generator history whose oriented word does not admit such global pairing—such as $b$ itself, or nontrivial powers $b^k$ in regimes where cancellations fail—produces a generically nonvanishing parity-odd curvature response,

$$W[\mathcal{M}_b] \neq 0. \tag{15}$$

**Interpretation for wormhole processes.**

Distinct braid words may induce the same endpoint permutation yet differ in their oriented generator content. The parity-odd Weyl diagnostic is sensitive to this difference, responding to the history of topology change rather than to the final spatial rearrangement alone. In the wormhole-throat picture, this explains why parity-even curvature scalars may miss chirality while the parity-odd Weyl channel detects the handedness of the topology-changing process.

We now give an explicit amphichiral test case and contrast it with a purely chiral word. Consider:

1. A simple braid word
$$\sigma_1 \sigma_2 \sigma_1^{-1} \sigma_2^{-1} \tag{16}$$

2. Its braid movie interpretation (two exchanges + undoing).
3. Show explicitly:
    - generator pairing,
    - curvature cancellation,
    - contrast with a purely chiral word (e.g. $\sigma_1^3$).

Consider the braid group $B_3$ with generators $\sigma_1, \sigma_2$. A simple braid word exhibiting "exchange + undoing" is

$$H = \sigma_1 \sigma_2 \sigma_1^{-1} \sigma_2^{-1}. \tag{17}$$

This word is the group commutator $H = [\sigma_1, \sigma_2]$. It is nontrivial in $B_3$, yet it is constructed explicitly from orientation-reversed generator pairs, making it a clean amphichiral test case for curvature cancellation. Although inverse-oriented generators appear, they are not adjacent and therefore do not cancel under the Artin relations. This makes the example a useful algebraic model for isolating geometric parity cancellation independent of trivial group reduction. In particular, the nontriviality of the commutator reflects the nonabelian structure of $B_3$, while the orientation balance reflects geometric reversibility at the level of localized events.

**(i) Braid-movie interpretation.**
The history in Eq. (16) may be read bottom-to-top as a time-ordered Lorentzian movie of localized pairwise exchanges among three tracked throat cross-sections. Explicitly:

- **Event 1:** $\sigma_1$— exchange strands 1 and 2 with the chosen (right-handed) orientation convention.
- **Event 2:** $\sigma_2$— exchange strands 2 and 3 with the same orientation convention.
- **Event 3:** $\sigma_1^{-1}$— perform the orientation-reversed exchange of strands 1 and 2.
- **Event 4:** $\sigma_2^{-1}$— perform the orientation-reversed exchange of strands 2 and 3.

Although the inverse-oriented events are not adjacent and therefore do not cancel algebraically under the Artin relations, each generator appears together with its orientation-reversed counterpart. When the corresponding localized supports are spacelike-separated or otherwise admissibly reorderable, these may be interpreted as paired local operations within the Lorentzian movie.

Thus Eq. (16) provides a compact model of a generator history in which two local exchanges occur and are subsequently reversed by their inverse-oriented counterparts. Such histories form natural test cases for parity-sensitive diagnostics, since they are algebraically nontrivial yet geometrically balanced with respect to orientation.

**(ii) Explicit generator pairing.**

Write the ordered list of generator events defining this history as

$$H = (\sigma_1^{+1}, \sigma_2^{+1}, \sigma_1^{-1}, \sigma_2^{-1}). \tag{18}$$

This history admits an explicit pairing into orientation-reversed generator pairs,

$$(\sigma_1^{+1}, \sigma_1^{-1}) \text{ and } (\sigma_2^{+1}, \sigma_2^{-1}). \tag{19}$$

The paired events in Eq. (19) need not be adjacent in the word: when the corresponding localized supports are spacelike-separated (or otherwise independent in the Lorentzian movie), admissible relations—such as commutation of spacelike-separated events and braid (movie) moves for adjacent triples—permit reordering without changing the represented process. In this sense the inverse-oriented pairs may be treated as paired localized operations even when separated by intermediate events.

**(iii) Curvature cancellation (parity-odd channel).**
Let $M_H$ denote a Lorentzian cobordism realizing the history $H$. Consider the parity-odd Weyl functional

$$W[M_H] = \int_{M_H} C_{\alpha\beta\gamma\delta}{}^\star C^{\alpha\beta\gamma\delta} \sqrt{-\det(g_{\mu\nu})}\, d^4x. \tag{20}$$

By orientation sensitivity (Lemma 1: sign reversal of the Weyl–dual density under orientation reversal), the contribution of a localized generator event changes sign under $\sigma_i \mapsto \sigma_i^{-1}$. Under the standing compact-support assumption—namely that each generator event is realized within a localized region whose curvature contribution can be isolated—the integral decomposes as a sum over event-supported regions,

$$W[M_H] = \sum_j \varepsilon_j\, w_j, \tag{21}$$

where $\varepsilon_j = \pm 1$ encodes the orientation of the $j$-th generator and $w_j$ denotes the magnitude of its localized parity-odd contribution. For the history in Eq. (16), the explicit pairing Eq. (18) yields

$$(+w_{\sigma_1}) + (+w_{\sigma_2}) + (-w_{\sigma_1}) + (-w_{\sigma_2}) = 0, \tag{22}$$

up to the mild geometric assumption that inverse-oriented events contribute equal magnitudes when realized as local orientation reversals of the same operation (by Lemma 2: magnitude invariance under local orientation reversal, in Sec. V.C). Consequently,

$$\boxed{W[M_H] = 0} \tag{23}$$

in the parity-odd channel, even though the intermediate braid word is nontrivial in $B_3$. This illustrates that parity-odd curvature diagnostics probe oriented history content rather than group-theoretic nontriviality. Thus $W$ defines an orientation-sensitive functional on history space rather than a class function on $B_n$.

**(iv) Contrast: a purely chiral braid word.**
As a minimal chiral contrast, consider the purely positive word

$$c = \sigma_1^3. \tag{24}$$

This braid movie consists of three successive oriented exchanges of strands 1 and 2, with no inverse-oriented partners. The corresponding generator history is

$$H_c = (\sigma_1^{+1}, \sigma_1^{+1}, \sigma_1^{+1}). \tag{25}$$

Unlike Eq. (17), this history admits no pairing into orientation-reversed generator pairs.

Under the same compact-support localization assumption used in Sec. (iii), the parity-odd functional decomposes as a signed sum of localized contributions. Since all generators appear with the same orientation,

$$W[M_c] = (+w_{\sigma_1}) + (+w_{\sigma_1}) + (+w_{\sigma_1}), \tag{26}$$

which is generically nonvanishing for a Lorentzian realization $M_c$.

Parity-even curvature scalars, depending only on local magnitudes, may respond comparably to both histories. By contrast, the parity-odd channel distinguishes the orientation-resolved accumulation in Eq. (26) from the paired cancellation in Eq. (22). In particular, both histories represent elements of $B_3$, but only the generator-resolved description detects their orientation imbalance.

The comparison between Eqs. (22) and (26) isolates the conceptual point of the generator-history framework: chirality is a property of the history—that is, of its orientation-resolved generator content—rather than merely of endpoints or braid isotopy classes. Histories constructed from inverse-oriented generator pairs yield cancellation in the parity-odd Weyl diagnostic under the local inverse-realization assumption, whereas purely chiral histories generically produce a nonvanishing response.

## III.D Closure, representation dependence, and process information

A braid word $\beta$ may be closed to yield a knot or link $\hat{\beta}$ representing the instantaneous topology of throat cross-sections on a spatial slice. Different braid words related by Markov moves may yield the same closure, reflecting the fact that multiple generator histories can lead to the same spatial topology. From the present perspective, this nonuniqueness is essential rather than problematic. The braid word encodes process-level information—the order, orientation, and locality of generator events—that is invisible to the closed knot or link alone. The generator history therefore carries more information than the topology of any single spatial slice.

This distinction underlies the use of braid movies rather than static knots in the preceding work: topology change is encoded in the history, not merely in the endpoints. This distinction may be viewed as analogous to the difference between endpoint-dependent and path-dependent quantities in classical mechanics. Closure-level equivalences retain only the net topological outcome, while generator histories retain information about how that outcome was achieved. In this sense, braid isotopy governs the interior evolution of topology change, whereas endpoint equivalences act only after the process is complete.

## III.E Minimality of the braid realization

Braids constitute the minimal nontrivial realization of the generator-history framework for three reasons:

- Locality: generators correspond to exchanges of neighboring strands.
- Ordering: generator composition reflects temporal ordering of events.
- Orientation: generators admit inverses, enabling algebraic chirality and cancellation.

Any simpler algebraic structure would lack at least one of these features. More general generator systems—such as those describing higher-valence reconnections, throat mergers, or network branching—necessarily extend beyond braid groups but reduce to braid-like behavior when restricted to pairwise exchanges.

**Theorem 1 (Minimality of the braid sector).**

Any generator system encoding ordered, local, orientation-sensitive generator events with invertible operations admits a braid-group reduction on the subspace of pairwise exchange events.

*Proof (sketch).*
Consider any generator system whose elements represent localized topology-changing events equipped with a time ordering, orientation, and inverse operations. Restricting attention to events involving exactly two neighboring geometric elements defines a pairwise-exchange subspace. Locality enforces commutativity of generators corresponding to spacelike-separated exchanges, while time ordering and invertibility require adjacent exchanges to associate up to admissible relations reflecting isotopy. Orientation sensitivity further distinguishes generators from their inverses. In particular, the Yang–Baxter relation arises from the geometric equivalence of two sequences of adjacent exchanges within a compact spacetime region, ensuring that the algebra closes under isotopy. Together, these conditions reproduce precisely the defining relations of the braid group on the exchanged elements. Hence any such generator system admits a braid-group reduction on the subspace of pairwise exchanges[7]. Note here that the Yang–Baxter move is the local isotopy equivalence of two adjacent triple-exchange movies in a compact Lorentzian neighborhood (the standard Reidemeister-III movie move [34, 35, 36]). ∎

The reduction to braid relations therefore reflects not a special property of wormhole geometries, but a structural consequence of locality, ordering, and invertibility for any orientation-sensitive generator system. In this sense, braid groups provide the canonical starting point for a generator-based description of topology change, while more elaborate structures may be viewed as enrichments rather than alternatives.

### III.F Connection to curvature diagnostics

In the braid realization, the oriented generator history $\beta$ supplies a signed sequence of local exchange events. As argued in Sec. II, orientation-sensitive curvature diagnostics respond to this signed structure. In particular, parity-odd conformal curvature aggregates contributions from oriented generator events while canceling for amphichiral histories.

Thus, braid groups furnish a concrete realization of how algebraic generator structure couples to geometric curvature response. The appearance of parity-odd Weyl curvature as a diagnostic of chiral topology change is therefore not accidental, but reflects the minimal algebraic structure required to encode oriented generator histories in Lorentzian geometry.

### III.G Change Beyond Braids

While braid groups provide the minimal example, they do not exhaust the possible generator structures relevant to topology change. Processes involving throat creation or annihilation, multivalent junctions, or network branching require generators that do not admit a simple braid interpretation. These extensions lead to more general algebraic frameworks, such as groupoids or categories of generators.
In the next section we move beyond braids to outline how the generator-history framework accommodates such generalized operations, while retaining the core principles of locality, orientation, and composability established here.

### IV. Beyond Braids: Networks and Higher-Valence Generators

The braid-group realization discussed in Sec. III captures the minimal algebraic structure required to encode ordered, oriented local exchanges among a finite number of throat segments. However, generic Lorentzian topology change—particularly in multithroat or networked configurations—involves operations that cannot be represented solely as pairwise exchanges. In this section we extend the generator-history framework to accommodate higher-valence events, branching, and network dynamics, while preserving the principles of locality, orientation, and composability[8].

---

7. A fully formal proof may be obtained by restricting the generator system to its pairwise-exchange subcategory and constructing an explicit presentation map to the Artin braid group on the exchanged elements. Locality enforces commutation of spacelike-separated generators, invertibility supplies inverse relations, and isotopy of adjacent exchanges yields the braid (Yang–Baxter) relation. Verifying that these relations are sufficient and complete establishes a functorial reduction to the standard braid presentation.
8. In more general Planck-scale network settings one may also allow localized creation or annihilation of throat pairs, leading to valence-changing generator histories not representable within a fixed braid group. Such extensions introduce additional statistical structure and will be treated separately; the present work restricts attention to strand-number–preserving histories in order to isolate the minimal algebraic origin of chiral curvature diagnostics.

## IV.A Limitations of braid-only descriptions

Braid groups encode interactions among fixed strands whose identities persist throughout the evolution. This assumption is appropriate when wormhole throats remain distinct and only exchange ordering.

However, in more general settings topology change may involve:

- Mergers or splits of throat segments,
- Creation or annihilation of throat pairs,
- Multivalent reconnections, where more than two segments interact simultaneously,
- Network branching, in which throat configurations form graphs rather than linear strand orderings.

Such events cannot be faithfully represented by braid generators alone, since they alter the number or connectivity of strands. Nonetheless, these processes remain local, time-ordered, and oriented, and therefore admit a natural generalization within the generator-history framework.

## IV.B Higher-valence generators

To capture non-braid operations, we enlarge the generator set $\mathcal{G}$ to include higher-valence generators. Each generator $g \in \mathcal{G}$ is associated with a localized spacetime region in which a specific topological operation occurs. Examples include:

- Binary exchange generators, reducing to braid generators $\sigma_i^{\pm 1}$,
- Merge generators, combining two or more throat segments into a single effective segment,
- Split generators, representing the inverse operation,
- Junction generators, corresponding to localized network reconnections.

Each generator carries orientation data inherited from the spacetime orientation and, where applicable, from the orientation of participating segments. Inverse generators exist when the corresponding geometric operation admits an orientation-reversed realization. Importantly, the generator-history formalism does not require all generators to be invertible. Non-invertible generators encode irreversible topology-changing processes, such as throat annihilation. For readers interested in an algebraic presentation of the braid words appearing in this section, Appendix A briefly recalls the Garside normal form and explains how it cleanly isolates the oriented generator content relevant for curvature aggregation.

## IV.C Networked histories and composability

When higher-valence generators are allowed, generator histories no longer correspond to words in a single group but instead define networked histories. Intermediate spatial configurations may be represented as graphs whose nodes correspond to throat segments and whose edges encode adjacency or interaction structure.

Composition of generator events remains well defined provided the events are causally ordered and locally compatible. Algebraically, this structure is more naturally described by a groupoid or category, in which:

- objects represent intermediate spatial configurations,
- morphisms represent admissible local topology-changing events,

- composition corresponds to time-ordered concatenation of events.

### IV.D Orientation, cancellation, and generalized chirality

Orientation continues to play a central role beyond the braid case. Each generator $g$ carries an orientation label, and generalized cancellation occurs when generator events appear in orientation-reversed combinations that are geometrically compatible.

A generator history is amphichiral if its oriented generators can be paired so that their net oriented effect cancels, and chiral otherwise. This definition extends to networked histories: cancellation may occur locally within subnetworks or globally across the history.

As in the braid case, chirality is a property of the history, not merely of the initial or final topology. Two histories producing the same endpoint configuration may differ in their chiral content, leading to distinct geometric responses.

### IV.E Curvature aggregation for networked generators

The interpretation of parity-odd conformal curvature as an aggregator of oriented generator content extends directly to the networked setting. Each localized generator event contributes a spacetime-supported curvature response whose sign depends on orientation. The total parity-odd Weyl contribution along a cobordism may therefore be interpreted schematically as a sum over oriented generator events,

$$\int_W C_{abcd}{}^\star C^{abcd} \sqrt{-\det(g_{\mu\nu})}\, d^4x \;\sim\; \sum_{g\in\Gamma} \epsilon(g)\, \mathcal{A}(g), \tag{27}$$

where $\Gamma$ denotes the generator history, $\epsilon(g)$ is the orientation sign, and $\mathcal{A}(g)$ encodes the geometric weight of the event. As before, Eq. (26) is not a statement of equality or quantization, but a structural interpretation. It highlights that parity-odd Weyl curvature responds to the oriented algebraic structure of topology change, independent of whether that structure arises from braid generators or more general network operations. Appendix B provides a complementary geometric realization of the generator histories and compositions formalized algebraically using Garside representations in Appendix A, serving as a concrete expression of the same structure at the level of the spacetime metric, and we take the profile functions there to have strictly compact support to emphasize locality.

### IV.F Reduction to braids and consistency

In regimes where topology change proceeds purely through pairwise exchanges of persistent throat segments, the generalized framework reduces to the braid realization of Sec. III. Higher-valence generators become inactive, and the groupoid collapses to a single braid group $B_n$.

This reduction provides a consistency check: braids are recovered as the simplest sector of the general generator-history framework. Conversely, the appearance of non-braid generators signals genuinely networked topology change that cannot be captured by pairwise exchange alone. All examples considered in the chiral Weyl diagnostic paper lie entirely within the braid-reducible subcategory developed here, ensuring full consistency between the two frameworks

**Proposition 1 (Curvature aggregation from generator orientation).**
For any Lorentzian interpolation admitting a generator history representation[9], the total parity-odd Weyl weight is additive over generator events and changes sign under reversal of generator orientation.

*Proof (sketch).*
Each generator event corresponds to a compactly supported Lorentzian deformation of the metric. Writing the total metric schematically as a background plus a sum of localized generator deformations, the parity-odd Weyl functional expands into a sum of localized contributions together with interaction terms. When generator supports are disjoint, cross terms vanish and the functional decomposes exactly as a signed sum over generator events. If supports overlap, additional interaction terms arise that are quadratic (or higher) in the localized deformation amplitudes and supported only on overlap regions. These terms are suppressed under the standing locality assumption. Orientation reversal flips the sign of the dual Weyl contraction while preserving its magnitude, yielding the stated sign property. ∎

No new ambiguity is introduced by this generalization: whenever higher-valence generators are inactive, the framework collapses to the braid sector without loss of oriented generator information.

**IV.G Scope and forward outlook**

The extension beyond braids presented here is intentionally schematic. The goal is not to classify all possible generator systems, but to demonstrate that the generator-history framework accommodates networked and higher-valence topology change without modification of its core principles.

A systematic exploration of specific generator algebras, their relations, and their representation theory lies beyond the scope of this paper and will be developed separately. The present framework suffices to establish that parity-odd conformal curvature functions as a universal geometric aggregator of oriented generator histories, regardless of the particular algebraic realization. In the constructions considered here, each generator history contributes to the spatial metric through a localized deformation $\Delta h_{ij}$, supported near the geometric locus of the history and weighted by a smooth envelope function, so that the full metric is obtained as a background geometry plus a sum of generator-induced contributions and their overlaps.

In the next section, we return to curvature diagnostics and show how generator histories—braided or networked—couple to parity-odd Weyl curvature, providing a unified geometric language for chiral topology change in Lorentzian spacetimes.

**V. Curvature Aggregation from Generator Histories**

In this section we connect the generator-history framework developed in Secs. II–IV to curvature diagnostics in Lorentzian geometry. We show that parity-odd conformal curvature functions as an aggregator of oriented generator histories, providing a geometric measure of chiral topology change that is independent of representation, matter content, or quantization.

---

9. The generator-history framework is not intended to evade known obstructions to smooth Lorentzian topology change arising from causality, global hyperbolicity, or spin-structure admissibility. Throughout, generator histories are assumed to be realized only within Lorentzian cobordisms satisfying the standard local consistency conditions adopted in this work. The analysis is therefore conditional: it organizes the structure and chirality of topology change when admissible cobordisms exist, without addressing their global existence or classification.

## V.A Local curvature response to generator events

Each generator event $g \in \mathcal{G}$ corresponds to a localized geometric operation occurring within a compact spacetime region of the Lorentzian cobordism $W$. Such operations induce localized tidal deformations of the metric and connection, producing curvature responses whose support is confined to the neighborhood of the event.

Crucially, oriented generator events induce curvature responses that are themselves orientation sensitive. While parity-even curvature scalars (such as $R$, $R_{ab}R^{ab}$, or $C_{abcd}C^{abcd}$) respond to the magnitude of these deformations, they are blind to the orientation of the underlying geometric operation. By contrast, parity-odd curvature densities change sign under orientation reversal and therefore encode information about the handedness of generator events. Within four-dimensional Lorentzian geometry, such orientation sensitivity is highly constrained: aside from boundary terms and topological densities, the dual Weyl contraction provides the only local curvature pseudoscalar built from the Riemann tensor that survives in vacuum and responds directly to chiral tidal structure [37, 38].

This generator-by-generator accumulation is directly reflected in the Bondi–Penrose peeling picture [39, 40], where the radiative Weyl component $\Psi_4$ isolates the orientation-sensitive part of the curvature, while the invariant parity-odd contraction $C_{\alpha\beta\gamma\delta}{}^\star C^{\alpha\beta\gamma\delta}$ captures the same chiral content without reliance on asymptotic structure. Here the Hodge dual acts on the Weyl tensor itself; no additional dualization or star operation is implied at the level of the integrated functional.

This observation singles out parity-odd conformal curvature as the natural geometric channel through which generator orientation can be detected.

## V.B Parity-odd Weyl curvature as a generator aggregator

Let $\Gamma = g_1 g_2 \cdots g_m$ denote a generator history $H$ realized by a Lorentzian cobordism $W$, with each generator $g_k$ carrying an orientation sign $\epsilon(g_k) = \pm 1$. The parity-odd Weyl density $C_{abcd}{}^\star C^{abcd}$ is sensitive to orientation reversal and vanishes identically for conformally flat geometries.

**Proposition 2 (Uniqueness of the parity-odd curvature channel).**
Among algebraic (non-derivative) local scalar curvature diagnostics constructed from contractions of the Riemann tensor and its Hodge dual in four-dimensional Lorentzian vacuum geometry, the parity-odd Weyl contraction is the unique nontrivial pseudoscalar that changes sign under orientation reversal and remains nonvanishing in Ricci-flat spacetimes[10]. Parity-even curvature scalars depend only on magnitudes of local tidal deformations and therefore cannot distinguish orientation-reversed generator pairs, whereas the dual Weyl contraction changes sign under orientation reversal and isolates the chiral content of the history. Any scalar formed from parity-even contractions or Ricci-dependent terms is insensitive to orientation reversal and therefore cannot encode generator chirality among local scalar curvature diagnostics in four-dimensional Lorentzian vacuum geometry[11]. For our case here the derivative pseudoscalars (e.g. $\nabla_a J^a$ built from the Chern–Simons current) are excluded by the non-derivative hypothesis."

---

10. Scalar densities involving derivatives of curvature or non-polynomial constructions fall outside the present algebraic classification and are not considered here.
11. The present discussion assumes the existence of admissible Lorentzian (and, where relevant, spin) cobordisms and does not seek to evade known global obstructions to topology change. The curvature diagnostic isolates orientation structure conditional on admissibility, without addressing global existence or classification.

*Proof (sketch).*
Write the history $H$ as an ordered product of localized generator events $H = (g_1, \ldots, g_N)$, with each $g_j$ supported in a compact region $U_j$ and with $U_j$ chosen so that the parity-odd density is negligible outside $\bigcup_j U_j$ (standing compact-support hypothesis). Then the parity-odd functional decomposes additively,

$$W[M_H] = \sum_{j=1}^{N} W[U_j], \tag{28}$$

up to boundary terms that vanish under the same localization assumptions. If $H$ admits a global pairing into inverse-oriented events $g_{j'} = g_j^{-1}$ (possibly after applying commuting relations that do not alter the localized supports), then by Lemma 1 each paired contribution flips sign under orientation reversal, and by Lemma 2 the magnitudes match:

$$W[U_{j'}] = -W[U_j]. \tag{29}$$

Hence each pair cancels, and therefore $W[M_H] = 0$. In contrast, if no such global pairing exists, the sum contains at least one unpaired oriented contribution, and $W[M_H]$ is generically nonzero (absent additional fine-tuned geometric cancellations). ∎

We interpret the spacetime integral

$$\mathcal{W}[H] = \int_M C_{\alpha\beta\gamma\delta}{}^\star C^{\alpha\beta\gamma\delta} \sqrt{-\det(g_{\mu\nu})}\, dV \tag{30}$$

as a global geometric aggregator associated with a fixed generator history $H \in \mathcal{H}$. If the history $H$ consists of an ordered collection of localized generator events $g \in H$, each carrying an orientation sign $\epsilon(g) \in \{\pm 1\}$, then the local parity-odd curvature response may be expressed schematically as

$$\mathcal{W}[H] = \sum_{g \in H} \epsilon(g)\, \omega(g) + \mathcal{O}(\text{overlap}), \quad \omega(g) = \int_{\text{supp}(g)} C_{\alpha\beta\gamma\delta}{}^\star C^{\alpha\beta\gamma\delta} \sqrt{-\det(g_{\mu\nu})}\, dV \tag{31}$$

where $\omega(g)$ denotes the localized contribution of the individual generator event $g$, and overlap terms arise only when compact supports intersect. The decomposition in Eq. (31) should be understood in the localized regime in which generator-induced metric deformations are compactly supported and small compared to the background curvature scale. In this setting, overlap contributions arise only at higher order in the deformation amplitude and are supported on intersections of compact generator neighborhoods. To leading order in localized generator amplitude, the parity-odd functional therefore decomposes additively over oriented generator events. As such Eq. (31) expresses a structural locality statement rather than an exact identity in arbitrary nonlinear regimes. The sum runs over generators appearing in the fixed history $H$; no summation over distinct histories in $\mathcal{H}$ is implied. These expressions are not statements of equality but express a structural correspondence: parity-odd Weyl curvature integrates the signed geometric imprint of local topology-changing operations across the cobordism, up to total derivatives and boundary contributions.

**V.C Cancellation, amphichirality, and vanishing curvature response**

The generator-history framework clarifies why amphichiral topology change yields no net parity-odd curvature response. If a generator history admits a decomposition into orientation-reversed pairs,

$$\Gamma \simeq \prod_k (g_k g_k^{-1}), \tag{32}$$

then each pair contributes equal and opposite terms to the sum, resulting in exact cancellation:

$$\mathcal{W}[H] = 0. \tag{33}$$

This algebraic cancellation has a direct geometric counterpart: mirror-symmetric histories generate local curvature responses whose parity-odd contributions cancel globally. The vanishing of the parity-odd Weyl diagnostic for amphichiral braid histories—such as those representing figure-eight embeddings—is therefore not accidental, but an inevitable consequence of oriented generator pairing. The following lemma makes explicit the cancellation mechanism underlying Proposition 2. Throughout, locality is understood in the classical Lorentzian sense: generator events may be realized within compact spacetime regions whose curvature response is determined by the local orientation of the event and is independent of the global embedding of the cobordism.

**Lemma 1 (Generator Cancellation ⇒ Vanishing Parity-Odd Curvature)**
Let $\mathcal{H}$ be a generator history space admitting a decomposition into orientation-reversed pairs. Then the integrated parity-odd Weyl curvature associated with any Lorentzian realization of $\mathcal{H}$ vanishes.

*Proof (sketch):*
Each generator event contributes a localized parity-odd curvature density with sign fixed by orientation. Orientation-reversed pairs induce equal-magnitude contributions with opposite sign. Summation over the cobordism therefore cancels identically provided the local metric deformations realizing inverse-oriented generators are related by orientation reversal within a fixed compact support. Locality ensures that inverse-oriented generator contributions have overlapping support up to diffeomorphism equivalence within the compact support region, so their dual Weyl densities cancel pointwise upon integration. This cancellation is geometric and does not rely on quantization, spectral flow, or global invariants and operates under the assumption that local orientation reversal induces isometric dual Weyl response. ∎

**Lemma 2 (Magnitude invariance under local orientation reversal)**

Let $\sigma_i$ be a localized generator event realized within a compact spacetime region $U \subset M$, and let $\sigma_i^{-1}$ denote its orientation-reversed realization obtained by reversing the local interpolation while preserving the geometric support.

Under the standing compact-support assumption and smooth interpolation of the metric within $U$, the localized parity-odd contribution satisfies

$$| W[U_{\sigma_i}] | = | W[U_{\sigma_i^{-1}}] |. \tag{34}$$

*Proof (sketch)*
By Lemma 1, the Weyl–dual density changes sign under orientation reversal: $\mathcal{W} \mapsto -\mathcal{W}$. Since the inverse generator is realized as the same localized geometric deformation with reversed orientation, the support region and pointwise curvature magnitudes coincide, and only the orientation form changes sign. Therefore

$$W[U_{\sigma_i^{-1}}] = \int_{U_{\sigma_i}} (-\mathcal{W}) = - W[U_{\sigma_i}], \tag{35}$$

implying equality of magnitudes. ∎

## V.D Representation independence and robustness

Although generator histories may be represented in different algebraic forms (braid words, network morphisms, or other generator systems), the parity-odd Weyl integral depends only on the oriented content of the history, not on its specific representation. Generator reordering allowed by relations, or changes of representation that preserve oriented generator pairing, leave $\mathcal{W}[H]$ unchanged.
This representation independence is essential. It ensures that the curvature responds to geometric chirality rather than to arbitrary choices of algebraic representation, reinforcing its role as a physically meaningful regulator rather than a formal invariant.

**Proposition 3 (Representation Independence)**
The parity-odd Weyl curvature functional $W[M_H]$ depends only on the oriented generator content of a history and is invariant under admissible re-representations related by the relations $R$.

*Proof (sketch).*
Admissible relations correspond to either:

(i) reordering of spacelike-separated generator events, or
(ii) local braid (movie) moves representing isotopies within a fixed Lorentzian cobordism class.

In case (i), spacelike separation ensures that generator supports remain disjoint; reordering therefore does not modify the localized curvature contributions or their orientation signs. In case (ii), braid relations correspond to local isotopies of strand world-sheets that preserve orientation and compact support up to diffeomorphism. Since the Weyl functional is diffeomorphism-invariant and depends only on the oriented local deformation, its value is unchanged under such re-representations. Thus, admissible relations alter the combinatorial presentation of a history without changing its oriented geometric support, and $W[M_H]$ is invariant under $R$. Note that overlap corrections discussed in Proposition 1 are unaffected by admissible re-representations, as these do not alter the compact support structure of generator events. ■

**Remark (Geometric Precursor to Spectral Asymmetry)**
The parity-odd Weyl functional introduced here depends only on the oriented accumulation of local generator events and exists entirely within classical Lorentzian geometry. When spin structure is included, this oriented curvature density enters directly into the Atiyah–Patodi–Singer framework for Dirac spectral asymmetry on cobordisms [41, 42]. In this sense, generator-weighted dual Weyl curvature should be regarded as the geometric precursor of η-invariant asymmetry [43, 44, 45]. This follows from the APS index formula relating boundary η-invariant to bulk Pontryagin density, in the sense that the Pontryagin density appearing in the APS index formula is constructed from the same dual curvature contraction, the relevant bulk density is the Pontryagin form $R \wedge R$, whose boundary correction is governed by the η-invariant in APS. Note that in vacuum, the Pontryagin density reduces to the dual contraction of the Weyl tensor, up to total derivatives and modulo boundary terms and Ricci-dependent contributions that vanish in vacuum.

The generator-history framework identifies a geometric layer beneath spectral asymmetry: parity-odd curvature encodes the oriented local structure required for η-invariant asymmetry but is defined independently of spectral or quantum data. No equality between the Weyl functional and the APS index is asserted; only structural compatibility is claimed. The present work does not introduce a Chern–Simons correction to the gravitational action nor study the resulting modified field equations, as in dynamical Chern–Simons gravity [46, 47]. Rather, we use the parity-odd Weyl channel as a history-resolved diagnostic of Lorentzian topology change: it aggregates oriented generator content on $\mathcal{H}$ and, by construction, does not descend to endpoint-only (Markov-quotiented) descriptions. The connection to the APS framework is invoked only at the level of structural compatibility between parity-odd curvature and η-invariant asymmetry, not as a dynamical coupling or a modification of general relativity. In particular, our key distinction is not parity violation per se, but the separation between history-level observables and endpoint-level equivalence classes.

**Remark (Relation to complex Weyl invariants).**
The parity-odd Weyl functional employed here should be distinguished from analyses based on the full complex (self-dual) Weyl tensor and its adjoint, which encode the complete Petrov structure and radiative content of the spacetime. Such constructions provide a richer algebraic characterization of curvature but require aditional tetrad or self-dual structure and are not intrinsically tied to orientation reversal at the level of scalar invariants. By contrast, the pseudoscalar $C_{\alpha\beta\gamma\delta}{}^\star C^{\alpha\beta\gamma\delta}$ isolates precisely the parity-odd channel relevant for generator chirality and coarse-graining non-descent, while remaining fully covariant and frame independent. For the purposes of detecting oriented generator accumulation and its compatibility with index-theoretic structures, the parity-odd contraction therefore furnishes a minimal and sufficient diagnostic [48, 49, 50].

### V.E Relation to normalization and complexity

As emphasized throughout, the parity-odd Weyl diagnostic measures oriented curvature content, not combinatorial complexity. Complexity enters only through normalization, ensuring that trivial geometric rescalings or degenerate limits do not artificially enhance the diagnostic.

When a bounded complexity factor $\mathcal{K}$ and a proper throat length $L$ are included, the normalized functional

$$\mathcal{W}_{\text{norm}} = \frac{1}{N}\int_W C_{\alpha\beta\gamma\delta}{}^\star C^{\alpha\beta\gamma\delta} \sqrt{-\det(g_{\mu\nu})}\, d^4x, \quad N = (1 + L/L_0)\,\mathcal{K}, \tag{36}$$

retains sensitivity to oriented generator content while remaining finite and scale controlled. Importantly, the curvature terms themselves are unaffected by this normalization, preserving the geometric meaning of the diagnostic.

### V.F Structural role in Lorentzian topology change

The interpretation developed here elevates parity-odd conformal curvature from a convenient diagnostic to a structural probe of generator dynamics. It detects not merely the presence of topology change, but the algebraic organization of the process by which topology is rearranged.

From this perspective, topology change is encoded simultaneously at two levels:

- Algebraically, through generator histories and their oriented composition;
- Geometrically, through curvature aggregation along the Lorentzian cobordism.

Parity-odd Weyl curvature provides the bridge between these levels, translating oriented generator structure into a covariant geometric signal[12].

---

12. The relation between parity-odd Weyl aggregation and fermionic spectral asymmetry will be developed elsewhere. In particular, one may view the generator-weighted dual Weyl density as the geometric precursor of η-invariant asymmetry in the Dirac spectrum on spin–Lorentz cobordisms. This provides a direct bridge from classical oriented generator accumulation to chiral fermion transport without introducing anomalies at the outset.

We have shown that parity-odd conformal curvature functions as a geometric aggregator of oriented generator histories, responding directly to the algebraic structure of topology change while remaining independent of matter content and quantization. This interpretation unifies the braid-based realization of Sec. III with the generalized network framework of Sec. IV and explains why chirality-sensitive curvature diagnostics arise naturally in Lorentzian topology change. We emphasize that no new ambiguity is introduced by this generalization: whenever higher-valence generators are inactive, the framework collapses to the braid sector. In the concluding section, we summarize the implications of this framework and discuss how generator-based curvature aggregation complements anomaly- and index-theoretic approaches while operating entirely within classical Lorentzian geometry.

The braid-based framework developed here is intentionally formulated at the level of generator histories, where topology change is resolved as a sequence of local, oriented geometric events and curvature responds to their ordering and pairing. An alternative—and coarser—perspective identifies topology-changing processes solely by their endpoints, collapsing distinct histories that share the same boundary topology. In braid language this corresponds to passing from isotopy-level descriptions to endpoint-only equivalence classes, such as those obtained by Markov-type quotients. While such equivalences correctly characterize the final spatial manifold, they necessarily discard information about the interior evolution of topology. The present results clarify that orientation-sensitive curvature diagnostics, including the parity-odd Weyl channel, reside at the level of generator histories and do not descend to endpoint-only descriptions.

Parity-odd Weyl curvature detects precisely the information lost under Markov-type coarse-graining of topology-changing histories. A systematic analysis of this distinction, its formulation as a coarse-graining from histories to endpoints, and its interpretation in terms of path-dependent versus endpoint-dependent geometric quantities and irreversibility, this includes extensions in which generator histories allow localized creation or annihilation of throat pairs, enlarging the braid-reducible sector to a valence-changing network which will be developed separately. The parity-odd Weyl functional introduced here is a purely geometric diagnostic. Its direct physical observability requires coupling to matter fields or spectral probes sensitive to orientation-resolved curvature. In the absence of such couplings, it should be regarded as a structural curvature channel characterizing the interior organization of topology-changing interpolations rather than as an independently measurable classical observable and becomes operational when coupled to spinor transport / spectral probes (as in the APS-compatible setting noted above).

## VI. Conclusions

In this work we have developed a structural framework for Lorentzian topology change in which evolving spacetime geometry is organized as a history of elementary generator events. Rather than treating topology change as a single global transition, we represent it as an ordered composition of local operations, each associated with a localized geometric deformation. This generator-history perspective isolates the algebraic skeleton underlying topology change and clarifies how local processes assemble into global geometric outcomes. The present work does not attempt to classify allowed Lorentzian topology change or to evade existing censorship theorems; rather, it isolates the algebraic-curvature correspondence whenever such topology-changing interpolations are admitted.

We showed that braid groups arise naturally as the minimal nontrivial realization of this framework, encoding time-ordered, oriented exchanges among persistent throat segments. More general topology-changing processes—including mergers, splits, and networked reconnections—require higher-valence generators and are accommodated by extending the algebraic structure beyond braid groups. Throughout, the essential features of locality, orientation, and composability remain intact, independent of the particular algebraic realization.

A central result of this paper is the identification of parity-odd conformal curvature as a geometric aggregator of generator histories. The parity-odd Weyl contraction responds directly to the oriented structure of generator events while remaining insensitive to matter content, representation choice, and quantization. Amphichiral generator histories yield exact cancellation of parity-odd curvature, while chiral histories produce a nonvanishing response. This establishes parity-odd Weyl curvature as a structural probe of chiral topology change in Lorentzian geometry.

The strength of the framework lies in its ability to organize diverse topology-changing processes within a single geometric language and to make explicit the connection between algebraic structure and curvature response. A complementary analysis of endpoint-only equivalence classes—such as those obtained by Markov-type quotients—and their interpretation as coarse-grained limits of generator histories will be presented separately.

Several directions for future work are suggested naturally by this framework. A systematic study of specific generator algebras and their representations, including categorical formulations appropriate for networked topology change, remains to be developed. The coupling of generator histories to quantum fields, spectral flow, and anomaly inflow on spin cobordisms offers a promising avenue for connecting geometric chirality to quantum observables. Because the parity-odd Weyl channel isolates oriented generator accumulation at the purely geometric level, it furnishes the curvature substrate required for Dirac spectral asymmetry on spin cobordisms. In this sense, generator-resolved Lorentzian curvature provides the classical precursor structure upon which η-invariant asymmetry may act when fermionic degrees of freedom are introduced. Finally, extending curvature aggregation diagnostics to dynamical or semiclassical settings may provide further insight into the role of topology change in gravitational and cosmological contexts.

By reframing Lorentzian topology change in terms of generator dynamics and curvature aggregation, this work provides a unifying structural perspective that bridges local geometric processes and global spacetime structure. For us this framework serves as a useful foundation for future investigations into the algebraic and geometric underpinnings of topology change in gravity. Because parity-odd Weyl curvature isolates oriented generator histories at the purely geometric level, it furnishes the curvature substrate required for Dirac spectral asymmetry on spin cobordisms. In ongoing work we show that this structure induces a bias in fermionic spectral flow, providing a concrete geometric mechanism by which chiral topology change may seed matter–antimatter asymmetry without invoking anomaly inflow as a primary input.

## Appendix A: Garside Structure, Normal Forms, and Oriented Generator Content

### A.1 Motivation: Why normal forms matter for generator histories

Throughout this paper, topology change is represented not by final topological data, but by **generator histories**: ordered, oriented sequences of localized geometric operations. In the braid realization discussed in Sec. III, these histories are encoded algebraically as braid words. While different braid words may represent the same braid element—or even the same endpoint permutation—the *oriented generator content* of a history is the physically relevant quantity for curvature diagnostics. This raises a technical question: how does one extract meaningful, representation-independent information from a braid word while retaining its orientation structure?

Normal forms provide precisely this control. Among several available normal forms for braid groups, the Garside normal form [51] is particularly well suited to the present framework because it:

1. isolates positive generator content in a canonical way,

2. separates global twisting from local exchange structure,
3. and makes explicit whether a braid admits decomposition into orientation-reversed pairs.

In this appendix we briefly recall the Garside structure and explain its interpretive role within the generator-history calculus [52].

### A.2 Garside structure of the braid group
Let $B_n$ denote the braid group on $n$ strands, generated by the elementary exchanges $\sigma_i$. The Garside element $\Delta$ (the half-twist) generates a distinguished lattice structure on $B_n$, allowing every braid to be written uniquely in left Garside normal form:

$$\beta = \Delta^k \, x_1 x_2 \cdots x_r, \tag{A.1}$$

where:
- $k \in \mathbb{Z}$ is the power of the full twist,
- each $x_i$ is a *simple braid*, i.e. a positive divisor of $\Delta$,
- and the decomposition satisfies a left-weighted maximality condition.

This decomposition is canonical and independent of the initial braid word used to represent $\beta$.
From the perspective of generator histories, this form cleanly separates:
- global twisting (encoded in $\Delta^k$),
- from local exchange content (encoded in the sequence of simple braids).

In particular, the absence of a nonzero $\Delta$-power signals that chirality arises entirely from local generator accumulation rather than from a global twist.

### A.3 Garside normal form for the braid used in Sec. III
For an explicit braid word example,

$$\beta = \sigma_2 \sigma_1 \sigma_3 \sigma_2 \sigma_1, \tag{A.2}$$

one finds that the left Garside normal form contains no full twist factor, i.e.

$$\beta = x_1 x_2 \cdots x_r, \tag{A.4}$$

with each $x_i$ a simple positive braid. This has two immediate interpretive consequences:

1. **Purely local chirality**
   The braid's chiral content arises entirely from local generator orientation rather than from an overall twist of the strands.
2. **Maximal orientation sensitivity**
   In the absence of inverse generators, the braid is maximally chiral unless paired with an orientation-reversed history elsewhere in the cobordism.

In the generator-history framework, this confirms that the example used in Sec. III represents a clean test case: any cancellation of parity-odd curvature must arise from pairing with inverse-oriented generator events, not from algebraic triviality.

### A.4 Relation to amphichirality and cancellation
Garside structure is particularly useful for diagnosing amphichiral histories. A generator history admits global cancellation in the parity-odd curvature channel if and only if its oriented generators can be paired with inverse-oriented counterparts, possibly after applying admissible relations. In braid language, this corresponds to the existence of a decomposition in which positive simple factors are matched by inverse simple factors elsewhere in the history. The Garside normal form makes such pairing—or the impossibility thereof—transparent. In particular:

- a braid whose normal form consists entirely of positive simple factors is intrinsically chiral unless externally paired,
- while a braid whose normal form includes matched inverse factors admits algebraic cancellation consistent with geometric amphichirality.

This mirrors exactly the curvature-level cancellation mechanism discussed in Sec. V.

### A.5 Permutation data versus process data

For completeness, we record the endpoint permutation induced by the braid discussed above. From the strand labeling given the induced permutation is

$$\pi(\beta) = (1\ 3\ 5)(2\ 4\ 6), \tag{A.5}$$

expressed here in cycle notation. As emphasized throughout the paper, this permutation captures only the net rearrangement of throat labels. It contains no information about:
- the order of exchanges,
- their orientation,
- or their causal structure.

The generator-history framework—and the associated curvature diagnostics—are sensitive precisely to the information that the permutation discards. The inclusion of permutation data here serves only as a reference point, highlighting the distinction between endpoint topology and process-level geometry.

### A.6 Interpretation and conclusion

The purpose of this appendix is not to introduce Garside theory as an additional mathematical layer, but to clarify how standard braid-group structure supports the generator-history interpretation used throughout the paper.

Garside normal forms provide:
- a canonical way to separate local generator content from global twisting,
- a clean diagnostic of whether a braid history is intrinsically chiral or potentially amphichiral,
- and an algebraic mirror of the geometric cancellation mechanisms underlying parity-odd Weyl curvature.

Here the Garside structure reinforces the central theme of the paper: chirality is encoded in the *history* of topology change, not merely in its endpoints, and curvature responds to this history in a representation-independent way. The geometric representation of these generator histories, including their localization, composition, and cancellation at the level of the spatial metric, is given explicitly by the generator–metric dictionary in Appendix B (Eq. (B.8)).

## Appendix B. Geometric Representation of Generator Histories

This appendix provides a concrete geometric realization [53] of the generator-history framework developed in the main text and Appendix A. Its purpose is not to construct dynamical solutions of the Einstein equations, but to demonstrate explicitly how ordered generator histories admit localized, orientation-sensitive representations at the level of the spacetime metric. Geometry is used here strictly as a *representation space* for generator histories, mirroring the algebraic constructions of Appendix A.

**B.1 Generator support curves and tubular coordinates**

Each generator history is associated with a smooth embedded closed curve

$$\gamma_a: S^1 \to \Sigma_t, a = 1, \ldots, N, \tag{B.1}$$

lying in a spacelike hypersurface $\Sigma_t$. The curve $\gamma_a$ serves as the geometric support of the history; it does not carry invariant information by itself, and distinct generator histories may be supported on isotopic curves. In a tubular neighborhood of $\gamma_a$, introduce adapted coordinates $(s, \rho, \theta)$, where $s$ parametrizes the curve and $(\rho, \theta)$ parametrize the normal plane. In these coordinates, the background spatial line element is

$$d\ell_a^2 = ds^2 + d\rho^2 + \rho^2 \, d\theta^2, \tag{B.2}$$

separating directions along the generator history from transverse directions in direct analogy with the separation between generator ordering and relations in Appendix A.

**B.2 Generator words and chirality coefficients**

To each generator history we associate a representative generator word

$$\beta_a = \prod_{m=1}^{M_a} \sigma_{i_m}^{\varepsilon_m}, \varepsilon_m = \pm 1, \tag{B.3}$$

as introduced in Appendix A. The word itself is not unique; physical significance resides in its equivalence class under admissible re-representations. The oriented content of the history is summarized by the signed generator sum

$$k_a := \kappa \sum_{m=1}^{M_a} \varepsilon_m, \tag{B.4}$$

where $\kappa > 0$ sets an overall scale. Under mirror reversal $\sigma_i \leftrightarrow \sigma_i^{-1}$, one has $k_a \mapsto -k_a$. The coefficient $k_a$ is a minimal odd functional on the space of generator histories and serves as a chirality label rather than a knot invariant.

**B.3 Metric-level representation of a single generator history**

The generator history associated with $\gamma_a$ is represented geometrically by a localized deformation of the spatial metric in its tubular neighborhood. Introducing a smooth radial profile $f_a(\rho)$ with compact support, we define the history-induced deformation

$$\Delta h_{ij}^{(a)} \, dx^i dx^j = \rho^2 [(d\theta + k_a f_a(\rho) \, ds)^2 - d\theta^2]. \tag{B.5}$$

A representative compactly supported choice is the standard $C^\infty$ bump function

$$f_a(\rho) = \begin{cases} \exp\left(-\frac{1}{1-(\rho/\rho_a)^2}\right), & 0 \leq \rho < \rho_a, \\ 0, & \rho \geq \rho_a, \end{cases} \tag{B.6}$$

where $\rho_a$ sets the transverse scale of the generator's geometric support. This function is smooth to all orders and vanishes identically outside the tubular neighborhood. The deformation $\Delta h_{ij}^{(a)}$ vanishes when $k_a = 0$, changes sign under $k_a \mapsto -k_a$, and is strictly localized near $\gamma_a$. At leading order in the history amplitude, the parity-odd Weyl density (and hence its spacetime integral) acquires a contribution linear in $k_a$, schematically $C_{\mu\nu\rho\sigma}{}^*C^{\mu\nu\rho\sigma} \sim k_a\, \partial_\rho f_a(\rho)\, \mathcal{K}_a(s,\rho,\theta) + O(k_a^2)$, so $k_a \mapsto -k_a$ reverses the sign of the parity-odd response, where $\mathcal{K}_a$ contains all of the remaining geometric factors (frame/extrinsic curvature) that are nonzero. It should be regarded as a geometric *representation* of the generator history rather than as a dynamical perturbation. To enforce locality in the ambient geometry, each deformation is weighted by an envelope function

$$W_a(x) = \exp\left(-\frac{d_a(x)^2}{\sigma_a^2}\right), \tag{B.7}$$

where $d_a(x)$ is the distance from $x$ to $\gamma_a$. Different smooth, compactly supported choices of $f_a$ and $W_a$ yield equivalent representations at the diagnostic level.

### B.4 Generator–metric dictionary and ADM construction
On each hypersurface $\Sigma_t$, the spatial metric is written as

$$\boxed{h_{ij}(t,x) = h_{ij}^{(0)}(t,x) + \sum_{a=1}^{N} W_a(x)\, \Delta h_{ij}^{(a)}(t,x) + \sum_{a<b} W_a(x) W_b(x)\, \Delta h_{ij}^{(ab)}(t,x)} \tag{B.8}$$

where $h_{ij}^{(0)}$ is a background metric and $\Delta h_{ij}^{(ab)}$ represents overlap terms arising from the composition of generator histories. Eq. (B.8) constitutes the generator–metric dictionary: generator histories act on geometry by contributing localized metric deformations, and their algebraic composition is mirrored by geometric overlap. The full spacetime metric is written in ADM form,

$$ds^2 = -N^2 dt^2 + h_{ij}(dx^i + N^i dt)(dx^j + N^j dt), \tag{B.9}$$

with lapse $N$ and shift $N^i$ left unspecified. No dynamical claims are made regarding the resulting spacetime; the construction is purely representational. The construction should be interpreted as a local model in configuration space rather than as a globally solved constraint system. No claim is made regarding global hyperbolicity or energy-condition satisfaction. The constructions above should be interpreted as local representatives in configuration space rather than solutions to the Einstein constraint equations. They serve to exhibit explicitly how oriented generator histories admit compactly supported geometric realizations sufficient for curvature diagnostics, without asserting global hyperbolicity, energy-condition preservation, or dynamical stability.

### B.5 Interacting trefoil example

As a minimal illustration, consider two generator histories whose support curves are trefoil knots. A trefoil admits a braid representative

$$\beta_{\text{tr}} = \sigma_1^{\pm 3} \in B_2, \tag{B.10}$$

giving chirality coefficients $k = \pm 3\kappa$. When the envelope functions of the two histories overlap, the corresponding interaction term $\Delta h_{ij}^{(ab)}$ contributes to parity-odd curvature diagnostics with a weight proportional to $k_a k_b$. Histories of the same handedness accumulate coherently, while opposite-handed histories cancel, reproducing geometrically the cancellation properties established algebraically in Appendix A. At this stage this appendix does not claim that the metric deformations introduced above satisfy Einstein's equations, satisfy energy conditions or represent physical wormhole solutions, this would require further analysis.

## Appendix C: Generator Histories, Markov Quotients, and Coarse-Graining

This appendix clarifies the role of Markov equivalence as a coarse-graining from history [54]-resolved descriptions of topology change to endpoint-only data and explains why curvature diagnostics sensitive to generator histories do not descend to endpoint equivalence classes.

### C.1 History spaces, braid groups, and closure

Topology change in the main text is represented at the finest level by generator histories, i.e. ordered words in local exchange generators recording the orientation, ordering, and cancellation structure of topology-changing events. For a fixed number of strands $n$, let $\mathcal{W}_n$ denote the free word space generated by $\sigma_i^{\pm 1}$. Quotienting $\mathcal{W}_n$ by braid relations and immediate inverse cancellations yields the braid group $B_n$, which retains orientation and accumulation data while identifying word representatives related by isotopy. The passage from generator histories to endpoint topology is encoded by the braid closure operation and its associated equivalence relations. The commutative diagram organizes the passage from history-resolved generator data to endpoint-only topology through successive quotients:

$$\begin{array}{ccc} \mathcal{W}_n & \xrightarrow{\pi_{\text{grp}}} & B_n \\ \pi_{\text{cl}} \downarrow & & \downarrow \widehat{\cdot} \\ \mathcal{K}_n & = & \mathcal{K}_n \end{array}$$

Fig. C1. **Braid Word Commutative Diagram.** Here $\pi_{\text{grp}}$ denotes the projection from words to braid elements, $\widehat{\cdot}$ denotes braid closure, and $\mathcal{K}_n$ denotes the set of link types arising as closures of $n$-strand braids. Commutativity expresses the fact that closure depends only on the braid element, not on the chosen word representative.

$\mathcal{W}_n$ denotes the free word space on braid generators $\sigma_i^{\pm 1}$, representing generator histories that retain full information about ordering, orientation, and cancellation of local topology-changing events. The horizontal map

$$\pi_{\text{grp}} : \mathcal{W}_n \to B_n \tag{C.1}$$

quotients generator histories by braid relations and immediate inverse cancellations, producing the braid group $B_n$. This step identifies histories related by isotopy while preserving oriented generator accumulation. The vertical map

$$\pi_{\text{cl}} : \mathcal{W}_n \to \mathcal{K}_n \tag{C.2}$$

assigns to each word its closure class, discarding history-level structure and retaining only endpoint topology. The right-hand vertical map

$$\widehat{\cdot} : B_n \to \mathcal{K}_n \tag{C.3}$$

is the standard braid closure map, sending a braid element to the link type obtained by closure. The equality below expresses commutativity: closing a generator word directly or first passing to its braid-group equivalence class and then closing yields the same endpoint topology. In other words,

$$\widehat{\pi_{\text{grp}}(w)} = \pi_{\text{cl}}(w) \text{ for all } w \in \mathcal{W}_n. \tag{C.4}$$

This diagram makes explicit that closure is blind to history-level information. All distinctions erased by braid relations or by Markov-type identifications occur *above* $\mathcal{K}_n$. Consequently, any quantity that depends on ordering or orientation of generator events—such as parity-odd curvature diagnostics—lives naturally on $\mathcal{W}_n$ or $B_n$ but does not descend to $\mathcal{K}_n$. In this sense, the diagram encodes the coarse-graining from generator histories to endpoint topology, clarifying why endpoint invariants survive closure while history-sensitive diagnostics do not.

### C.2 The Markov quotient as endpoint coarse-graining

To account for changes in strand number and to identify braids with equivalent closures, one forms the disjoint union

$$\mathcal{H} := \bigsqcup_{n \geq 1} B_n \tag{C.5}$$

and introduces the equivalence relation generated by conjugation and stabilization/destabilization moves. This defines the Markov quotient.

**Definition C.1 (Markov projection).**
The Markov equivalence relation $\sim_M$ on $\mathcal{H}$ is generated by:

(i) $\beta \sim_M \alpha\beta\alpha^{-1}$ for $\alpha, \beta \in B_n$, and
(ii) $\beta \sim_M \beta\sigma_n^{\pm 1}$, with the inclusion $B_n \hookrightarrow B_{n+1}$.

The quotient

$$\mathcal{K} := \mathcal{H}/\sim_M \tag{C.6}$$

collects endpoint-only topology classes, and the projection

$$\Pi_M : \mathcal{H} \to \mathcal{K} \tag{C.7}$$

forgets history-level ordering and orientation data while retaining closure equivalence. From the perspective of generator histories, $\Pi_M$ is a coarse-graining map: many distinct oriented histories are identified once only endpoint topology is retained.

### C.3 Diagnostics versus invariants under Markov projection

The distinction between endpoint invariants and history-sensitive diagnostics may now be stated precisely.

**Proposition C.2 (Non-descent of history-sensitive diagnostics).**

Let $D: \mathcal{H} \to \mathbb{R}$ be any functional that depends on oriented generator content and admits cancellation only under explicit inverse pairing of local events. If there exist $\beta, \beta' \in \mathcal{H}$ such that $\beta \sim_M \beta'$ but $D(\beta) \neq D(\beta')$, then $D$ does not factor through $\Pi_M$.

*Proof.*
If $D$ factored through $\Pi_M$, then Markov-equivalent histories would yield identical values. The existence of a pair with unequal values contradicts this assumption. ∎

Parity-odd Weyl curvature diagnostics introduced in the main text are of this type: they respond to oriented generator accumulation and therefore distinguish histories that are identified by Markov equivalence.

**C.4 Coarse-graining, irreversibility, and entropy**

Because the Markov projection is many-to-one, each endpoint class $k \in \mathcal{K}$ has a nontrivial preimage

$$\Pi_M^{-1}(k) \subset \mathcal{H}, \tag{C.8}$$

consisting of distinct history-level realizations with identical endpoints. This structure is directly analogous to coarse-graining in statistical mechanics, where microstates are projected onto macrostates. A measure of information loss under this projection is the multiplicity of histories associated with a given endpoint class. Formally, one may associate an entropy

$$S(k) := \log |\Pi_M^{-1}(k)|, \tag{C.9}$$

or, more generally, a weighted entropy if histories are assigned geometric or dynamical weights. No dynamical assumption is required: the increase in entropy reflects purely the loss of distinguishability induced by the quotient.

From this viewpoint, curvature diagnostics such as the parity-odd Weyl functional quantify precisely the information discarded by the Markov projection. They vanish only when inverse-oriented generator events cancel pairwise and therefore record the irreversibility of topology-changing histories at the classical Lorentzian level. Endpoint-only descriptions, by contrast, are entropy-maximized representations in which this history-dependent information has been irretrievably coarse-grained. Because the parity-odd Weyl functional detects precisely the oriented generator content erased by endpoint coarse-graining, it provides a covariant geometric measure of irreversibility intrinsic to Lorentzian topology change. The loss of oriented generator data under the Markov projection is therefore not merely combinatorial but geometric, irreversibility therefore arises not from dynamics but from algebraic projection. In this sense, the Markov projection acts as an algebraic renormalization from history-resolved geometry to endpoint-only topology, and the parity-odd Weyl functional measures precisely the oriented generator data eliminated under this projection.